\documentclass[12pt,a4paper]{conference}

\usepackage{fancyhdr}
\usepackage{graphicx,amsmath,amssymb,cite}
\usepackage{multind}
\makeindex{author} \makeindex{subject}

\newcommand{\beq}{\begin{equation}}
\newcommand{\eeq}[1]{\label{#1}\end{equation}}
\newcommand{\eeqn}{\end{equation}}

\newcommand{\beqa}{\begin{eqnarray}}
\newcommand{\eeqa}[1]{\label{#1}\end{eqnarray}}
\newcommand{\eeqan}{\end{eqnarray}}

\let\bar=\overbar

\newcommand{\Dslash}{\not{\hbox{\kern-4pt $D$}}}
\newcommand{\dslash}{\not{\hbox{\kern-2pt $\del$}}}

\newcommand{\msb}{{\bar{\ssstyle M \kern -1pt S}}}

\newcommand{\lsim} {\buildrel < \over {_\sim}}

\begin{document}
%%%%%%%%%%%%%%%%%%%%%%%%%%%%%%%%%%%%%%%%%%%%%%%%%%%%%%%%%%%%%%%%%%%%%%%

\Chapter{Electromagnetic reactions on light nuclei using chiral
  effective theory}
        {Electromagnetic reactions on light nuclei using $\chi$ET}
	{D.~R.~Phillips}
\vspace{-6 cm}\includegraphics[width=6 cm]{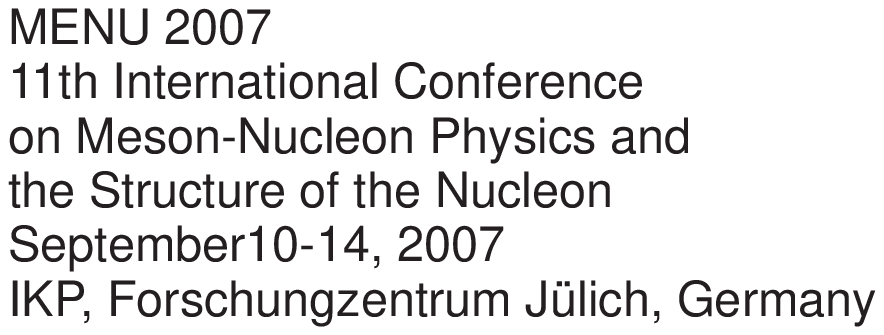}
%\bigskip\bigskip
\vspace{4 cm}

\addcontentsline{toc}{chapter}{{\it D.~R.~Phillips}} \label{authorStart}
%%%%%%%%%%%%%%%%%%%%%%%%%%%% NEW SWITCHES %%%%%%%%%%%%%%%%%%%%%%%%%%%%%%

\begin{raggedright}

{\it D. R. Phillips}
\index{author}{Phillips, D.R.}\\
Department of Physics and Astronomy\\
Ohio University\\
Athens, OH 45701\\
U. S. A.
\bigskip\bigskip

\end{raggedright}

\begin{center}
\textbf{Abstract}
\end{center}
I describe the use of chiral effective theory ($\chi$ET) to
compute electromagnetic reactions in two- and three-nucleon systems. I
first explain how chiral perturbation
theory can be extended to the few-nucleon sector. I then explain
the predictions of the resulting $\chi$ET for electron-deuteron
scattering, and how they will be tested by forthcoming data from
BLAST. I conclude by displaying predictions for elastic Compton
scattering from deuterium and Helium-3 nuclei. These computations, in
concert with future data from MAX-Lab and HI$\gamma$S, should give
significant new information on neutron polarizabilities, and hence
yield insight into the structure of the nucleon.

\section{Introduction}
Chiral perturbation theory ($\chi$PT) is the effective theory of the
strong interaction at low energies. In $\chi$PT quantum-mechanical
amplitudes for the interaction of pions and photons with each other
and with nucleons are expanded in powers of the small parameter $P$, where
$P \equiv \frac{p,m_\pi}{\Lambda_{\chi SB}}$.
The scale $\Lambda_{\chi SB} \sim m_\rho, 4 \pi f_\pi$ in the
meson sector, but is somewhat lower for reactions involving baryons
unless additional degrees of freedom (in particular the Delta(1232))
are included explicitly in the theory.

The amplitudes we seek are computed using the technology of effective
field theory (EFT), in which the field-theoretic Lagrangian is organized in
an expansion in powers of $P$ and loop calculations are then also
organized via the same hierarchy. 
Computing the $\chi$PT result for a given
process at a fixed order in $P$ is simply a matter of writing down the
Lagrangian up to that order and computing all the pertinent diagrams.
An introduction to, and explicit examples of, this strategy was given
in Prof.~Gasser's talk at this meeting~\cite{Gasser}.

This approach has had much success in treating $\pi \pi$ and $\pi N$
interactions at energies below $\Lambda_{\chi \rm SB}$ (see
Ref.~\cite{Bernard2006} for a recent review). However, an obvious
problem arises when we attempt to extend it to light nuclei: a
perturbative expansion of amplitudes is not adequate to describe bound
states. In 1990 Weinberg proposed that the fact that the nucleon mass,
$M \sim \Lambda_{\chi \rm SB}$, mandates resummation of diagrams with
$NN$ intermediate states, and so, when computing $NN \rightarrow NN$,
$\chi$PT should be applied not to the $NN$ amplitude, but to the $NN$
potential $V$~\cite{Weinberg}. In such an expansion the leading-order
(LO) $NN$ potential, $V$, is:
\begin{equation}
\langle {\bf p}'|V|{\bf p} \rangle=-\frac{g_A^2}{4 f_\pi^2} \tau_1 \cdot \tau_2
\frac{\sigma_1 \cdot ({\bf p}' - {\bf p})  \sigma_2 \cdot ({\bf p}' -
  {\bf p})}{({\bf p}' - {\bf p})^2 + m_\pi^2}  + C,
\label{eq:LOV}
\end{equation}
where $g_A$ and $f_\pi$ are the nucleon's axial charge and the
pion-decay constant, and the constant $C$ is
not determined by chiral symmetry and must be obtained from $NN$ data.
$V$ is then inserted into the Schr\"odinger equation
\begin{equation}
\left(\frac{\hat{\bf p}^2}{M} + V\right)|\psi \rangle=E|\psi \rangle,
\label{eq:SE}
\end{equation}
to generate bound and scattering states of two, or more, nucleons.
This strategy, which has come to be known as ``chiral effective
theory'' ($\chi$ET) produces a quantum-mechanical description of light nuclei,
in which the potential $V$ (and other operators too) have a
systematic chiral expansion and a rigorous connection with the chiral
symmetry of QCD and the pattern of its breaking.

Since the potential $V$ is singular a cutoff, $\Lambda$, must be
introduced. The constant $C$ is then a function of $\Lambda$. The
question arises as to whether this will be sufficient to renormalize
the $NN$ amplitude obtained by iterating $V$, i.e. whether there is
significant residual $\Lambda$-dependence in $NN$ observables after the
value of $C$ is adjusted to reproduce the very-low-energy $NN$ data.

There has been much debate on this point over the past 10 years, but
it has now been shown that a single constant $C$ is sufficient to
renormalize the $NN$ problem in the ${}^3$S$_1$--${}^3$D$_1$ channel
at LO~\cite{Beane2002,Pavon2005,Birse2006}. Moreover, these papers
argue that it is necessary to solve the Schr\"odinger equation with
the LO chiral potential precisely because that potential is not weak.
In contrast to the $A=0$ and $A=1$ sectors a perturbative expansion
for the $NN$ interaction mediated by pions only converges for $p \lsim
m_\pi$: the one-pion exchange part of $V$ is strongly
attractive---singular even---in the ${}^3$S$_1$--${}^3$D$_1$ channel.

More recently it has been pointed out that there are channels of
higher angular momentum where the LO potential (\ref{eq:LOV}) is also
attractive, but where the constant $C$ is not
operative~\cite{NTvK05}. Consequently it is impossible to generate
$\Lambda$-independent predictions in those channels (e.g. ${}^3$P$_0$)
over a wide cutoff range. How wide a $\Lambda$ range should be
employed is still debated~\cite{EM06}. Ultimately
renormalization-group techniques would seem the best way to determine
what operators must be included to renormalize $\chi$ET to a given
level of accuracy~\cite{Birse2006}.  This is an ongoing discussion.

But its ultimate resolution should not have a significant impact on
the results I present here, which are predominantly for deuterium,
where this is a solved problem (at least at LO). In Sec.~\ref{sec-ed}
I show results for deuteron electromagnetic form factors as a function
of the cutoff $\Lambda$ and demonstrate that cutoff artifacts indeed
disappear as $\Lambda \rightarrow \infty$. I also describe how a
chiral expansion for the deuteron charge operator generates precision
predictions for the ratio $G_C/G_Q$ that was recently measured at
BLAST. In Sec.~\ref{sec-Compton} I summarize $\chi$ET calculations of
elastic photon scattering from deuterium and Helium-3 nuclei. And in
Section~\ref{sec-other} I provide a brief summary of other reactions
involving light nuclei that have been successfully described in
$\chi$ET.

\section{Electron-deuteron scattering in $\chi$ET}

\label{sec-ed}

In Ref.~\cite{NTvK05} Eq.~(\ref{eq:SE}) was solved for the potential
(\ref{eq:LOV}) in momentum space for $\Lambda= 0.4$--$4$ GeV. In
Ref.~\cite{Pavon2005} the same problem was solved in co-ordinate space
by converting $C$ into a boundary condition on the wave function as $r
\rightarrow 0$~\cite{Pavon2005}. We now present results for deuteron
electromagnetic form factors that show that the
latter wave function can be regarded as the $\Lambda
\rightarrow \infty$ limit of the Fourier transform of the
momentum-space wave functions~\cite{Pavon2007}.

%These deuteron wave functions are then used to generate
%electromagnetic deuteron form factors according to the standard
%formulae (see e.g. Ref.~\cite{Ph07}).
%formulae:
%\begin{eqnarray}
%\qquad G_C=\frac{1}{3 |e|} \left(\left \langle 1\left|J^0\right|1 \right \rangle + 
%\left \langle 0\left|J^0\right|0 \right \rangle + \left \langle -1\left|J^0\right|-1 \right \rangle
%\right),\\ 
%G_Q=\frac{1}{2 |e| \eta M_d^2} \left(\left \langle 0\left|J^0\right|0 \right \rangle - \left \langle 1\left|J^0\right|1 \right \rangle\right), \quad
%G_M=-\frac{1}{\sqrt{2 \eta}|e|}
%\left \langle1\left|J^+\right|0\right \rangle,
%\end{eqnarray}
%where we have labeled these (non-relativistic) deuteron states by the
%projection of the deuteron spin along the direction of the momentum
%transfer ${\bf q}$ and $\eta \equiv |{\bf q}|^2/(4 M_d^2)$. $G_C$,
%$G_Q$, and $G_M$ are related to the experimentally-measured $A$, $B$,
%and $T_{20}$ in the usual way, with $T_{20}$ being primarily sensitive
%to $G_Q/G_C$ and $B$ depending only to $G_M$. Here we will compare
The deuteron charge and quadrupole form factors $G_C$ and $G_Q$
involve matrix elements of the (Breit-frame) deuteron charge operator
$J_0$ between these wave functions (for formulae see,
e.g. Refs.~\cite{vOG,Ph07}). Here we will compare $\chi$ET predictions for
$G_C$ with extractions from data for the deuteron
structure function $A$ and the tensor-polarization observable
$T_{20}$~\cite{Ab00B}. For this purpose we use the deuteron current operator
\begin{eqnarray}
\langle {\bf p}'|J_0({\bf q})|{\bf p}\rangle&=&
|e|  \delta^{(3)}(p' - p - q/2) G_E^{(s)}(Q^2) \label{eq:J0},
\end{eqnarray}
with $G_E^{(s)}$ the nucleon's isoscalar electric form factor. This is
the result for $J_0$ up to corrections suppressed by $P^3$
(apart from some small effects that have coefficients $\sim
1/M^2$). The strict LO $\chi$ET result for deuteron form factors is
found by taking $G_E^{(s)}=1$. However, here we wish to test
$\chi$ET's predictions for deuteron structure, so we adopt
``experimental'' data for $G_E^{(s)}$ (the parameterization of
Belushkin {\it et al.}~\cite{BHM07}) and compute $G_C$. Up to the
order we work to this is equivalent to computing the ratio
$G_C/G_E^{(s)}$ in $\chi$ET.

\begin{figure}[ht]
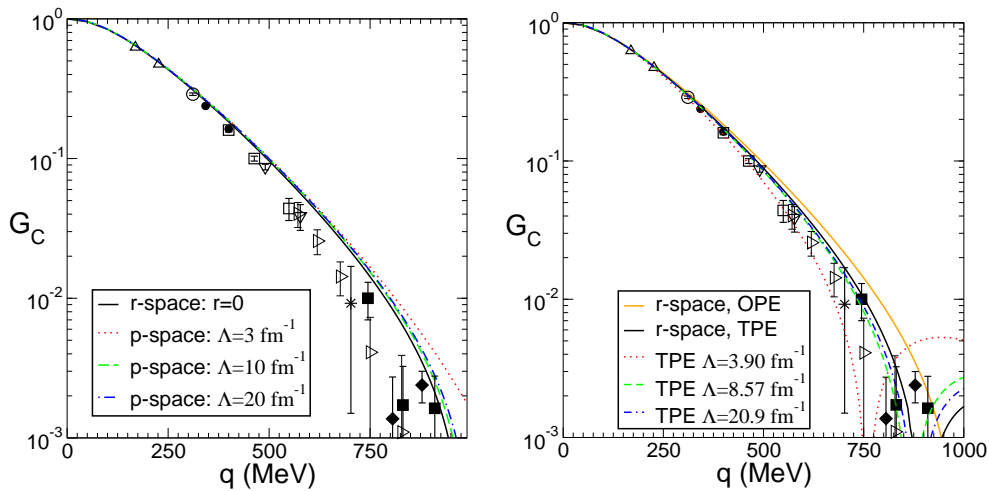

\vspace*{-0.2cm}
\begin{center}
\hskip -2.4in
\includegraphics[width=2.39in]{GC.LO.eps}
\vskip -2.52in
\hspace*{2.8in}
\parbox{2.5in}{\includegraphics[width=2.5in]{GC.TPEOPE.eps}}
\caption{Predictions for $G_C$ with LO wave functions (left panel) and
  wave functions including two-pion exchange (right panel).  In each
  case results for four different regulators are shown.  Data are
  taken from Ref.~\cite{Ab00B}.}
\label{fig-GC}
\end{center}
\vspace*{-0.5cm}
\end{figure}

The results are shown in the left panel of Fig.~\ref{fig-GC}.  Here
the value of $C$ is adjusted to ensure that the deuteron binding
energy is reproduced, but the calculation contains no other free
parameters.  We observe that as $\Lambda \rightarrow \infty$ the
momentum-space wave functions produce a $G_C$ that converges to a
definite result (although asymptopia is not
reached in $G_C$ until $\Lambda \approx 10~{\rm
  fm}^{-1}$). This result is consistent with that
found using the co-ordinate space approach of Ref.~\cite{Pavon2005}.  The
agreement with experimental data at low-$|{\bf q}|$ is quite good, but
the LO wave functions predict a minimum in $G_C$ at too large a
$Q^2$. These trends are even more pronounced in $G_Q$ (not shown),
where we are within a few per cent of the asymptotic result at
$\Lambda=4~{\rm fm}^{-1}$, and the agreement with experimental data is
excellent to a surprisingly large value of $|{\bf q}|$.

To go beyond LO we must consider corrections to the $NN$
potential $V$, and to the charge operator $J_0$. When the
two-pion-exchange mechanisms that define $V$ up to ${\cal
  O}(P^3)$~\cite{chiTPE} are included the renormalization becomes a
little more complicated since there are three undetermined parameters
summarizing ${}^3$S$_1$--${}^3$D$_1$ physics at scales $>
\Lambda_{\chi SB}$. Here these constants are adjusted to reproduce the
deuteron properties $B=2.22457$ MeV, $A_S=0.885~{\rm fm}^{-1/2}$ and
$\eta=0.0256$~\cite{Pavon2007}. Once again we see that convergence to the
$\Lambda \rightarrow \infty$ result is somewhat slow, but a smooth
$\Lambda \rightarrow \infty$ limit does exist. We also see that the
two-pion-exchange corrections to $V$ result in only a small shift in
$G_C$ in the range $|{\bf q}| < 800$ MeV. This lends credence to the
idea that these corrections could be treated in perturbation
theory---at least in the ${}^3$S$_1$--${}^3$D$_1$ channel. It is also
significant that these corrections shift the $G_C$ minimum to the left
as compared to the LO result, thereby improving the agreement with
experiment.  This suggests that electron-deuteron data provide
evidence for the presence of two-pion-exchange pieces in the $\chi$ET potential
$V$.

The existence of the minimum in $G_C$ provides an opportunity to
examine the impact on $G_C$ of the meson-nucleon dynamics that enters the
chiral $NN$ potential. In particular, different choices of
the $\pi$N LECs $c_i$ that enter $V$
produce minima in somewhat different
locations~\cite{Pavon2007}. However, it is difficult to draw any real
conclusion as regards the preference of existing $G_C$ data for a
particular set of $c_i$'s, since the ${\cal O}(P^3)$ corrections to $J_0$
that were not included in Eq.~(\ref{eq:J0}) have an impact on the
position of the minimum of $G_C$ that is at least as large as the
effect of choosing different sets of $c_i$'s. 

We close this section by pointing out that this type of analysis, when
carried out to higher orders in $\chi$ET, results in a precise
prediction for the ratio of deuteron form factors $G_C/G_Q$ in the
kinematic range of forthcoming data from BLAST. In Ref.~\cite{Ph07}
all contributions to $J_0$ (including two-body effects) up to order
$P^3$ relative to leading were computed. A variety of wave functions
that included all the two-pion-exchange effects up to ${\cal O}(P^3)$
were also employed.  Significant sensitivity to short-distance $NN$
physics was found in the resulting deuteron quadrupole moment $Q_d$:
it varied by 2\% when the cutoff in the $NN$ system was changed by
$\sim 100$\%. Intriguingly, this is roughly the magnitude of the
discrepancy between the $Q_d$ predicted at ${\cal O}(eP^3)$ and the
experimental value $Q_d=0.2859(3)~{\rm fm}^2$. This encouraged us to
include in our analysis a short-distance operator that represents the
contribution of modes above $\Lambda_{\chi \rm SB}$ to $G_Q$.  This
operator has much slower $Q^2$-dependence than the one-body mechanisms
that give the LO contribution to $G_Q$, so we can constrain its impact
by demanding that its coefficient is such that the experimental $Q_d$
is reproduced. This vitiates our ability to predict $G_Q$ at $Q^2=0$,
but we can still predict the $Q^2$-dependence of $G_Q$. The
prediction's remaining theoretical uncertainty---which comes from the
$Q^2$-dependence of short-distance $NN$ physics---is small, being only
3\% at $|{\bf q}|=2~{\rm fm}^{-1}$.  It is important to note that the
$\chi$ET predictions for $G_C/G_Q$~\cite{Ph07} have a rather different
$Q^2$-dependence to those obtained in potential models (see, e. g.,
Fig.~11 of Ref.~\cite{vOG}), and so the BLAST data will provide a
significant test of this approach to deuteron electromagnetic
structure.

\section{Compton scattering on $A=2$ and $3$ nuclei}

\label{sec-Compton}

Now we turn our attention to Compton scattering from the deuteron. The
first calculation of this process in $\chi$ET computed the $\gamma$d
amplitude
\begin{equation}
{\cal A} \equiv \langle \psi_d|\hat{O}|\psi_d \rangle
\end{equation}
by considering the operator $\hat{O}$ up to ${\cal O}(e^2 P)$ [NLO] and using
a variety of phenomenological deuteron wave
functions~\cite{Beane99}. At this order $\chi$ET makes a prediction for
${\cal A}$, and hence for the $\gamma$d differential cross section (dcs)~\cite{Beane99}, as
well as single- and double-polarization observables~\cite{CP04}.

\begin{figure}[ht]
\begin{center}
\hskip -2.5in
\includegraphics[width=2.8in]{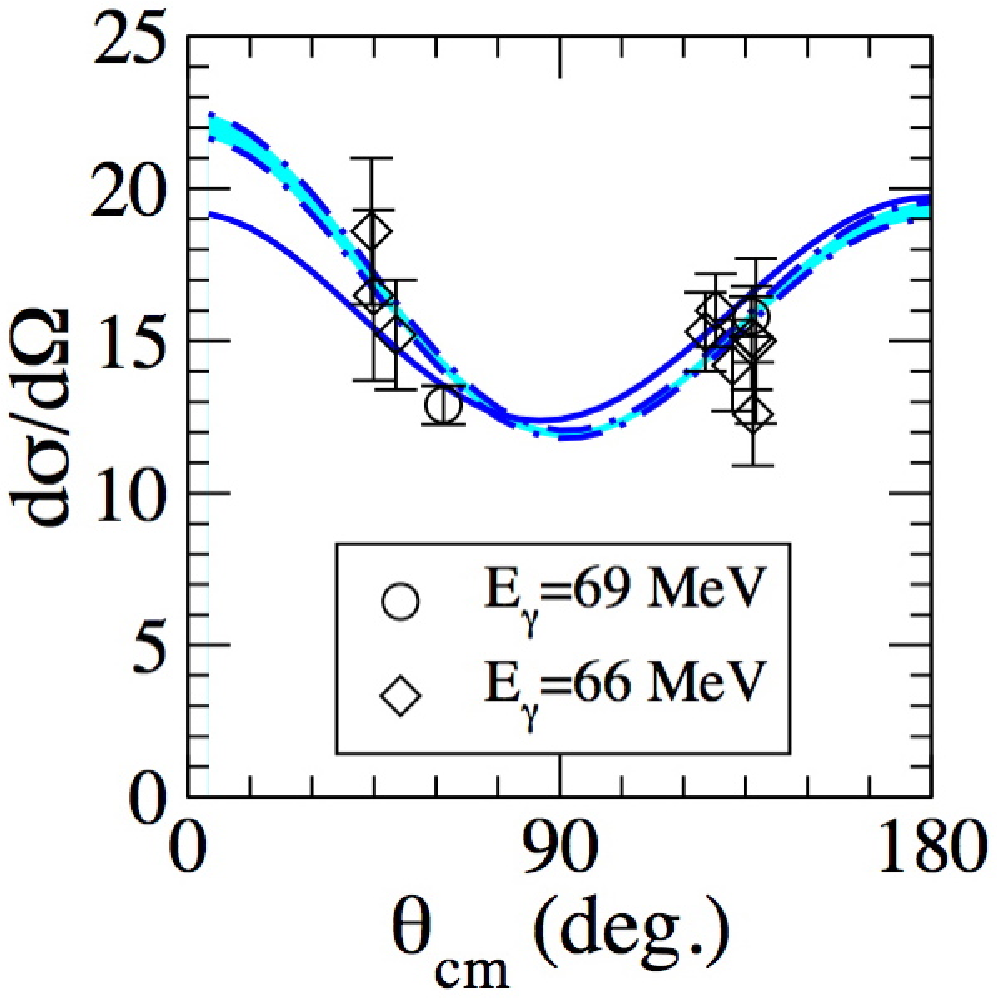}
\vskip -2.73in
\hspace*{2.8in}
\parbox{2.8in}{\includegraphics[width=2.48in]{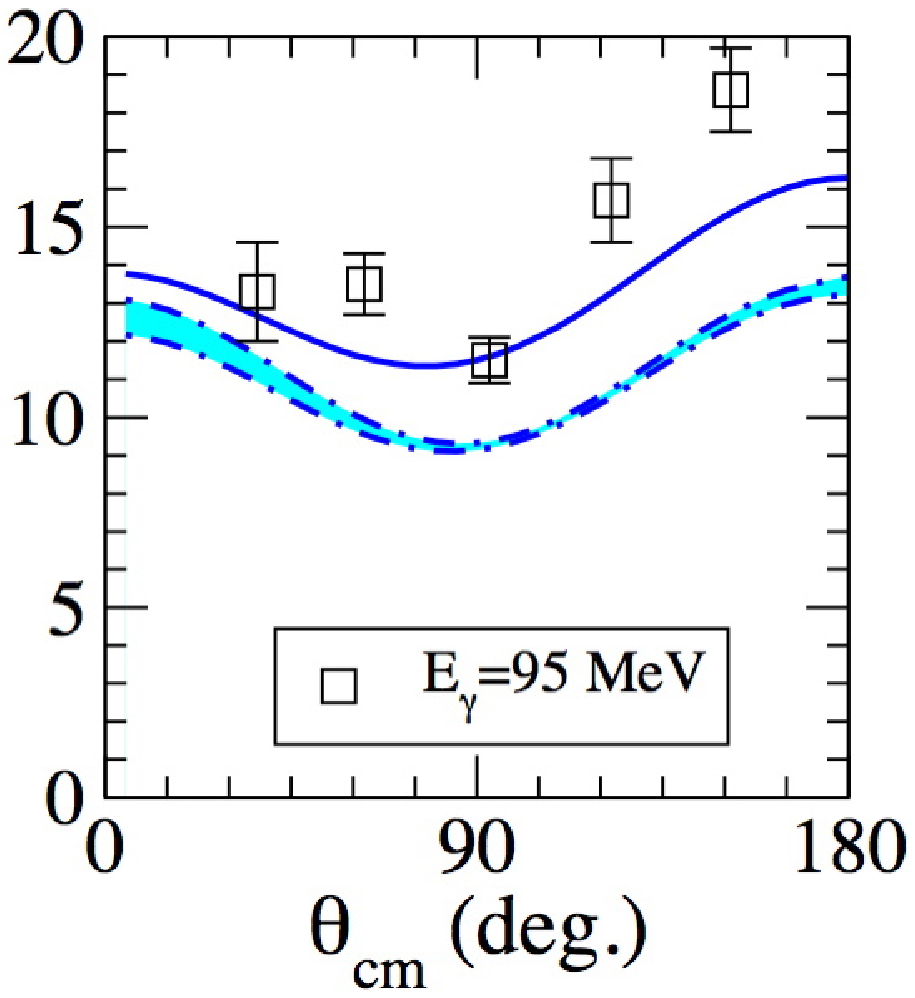}}
\caption{Centre-of-mass frame $\gamma$d dcs at $\omega=67$ 
  and 94.5 MeV respectively. The dot-dashed line is the prediction
  at ${\cal O}(e^2 P)$, with the band an estimate of the uncertainty
  due to short-distance $NN$ physics. The solid line is a fit at ${\cal
    O}(e^2 P^2)$. Adapted from Ref.~\cite{Beane05}, which includes references
  to data.}
\label{fig:gammad}
\end{center}
\vspace*{-0.5cm}
\end{figure}

For photon energies $\omega$ such that $\frac{m_\pi^2}{M} \ll \omega
\sim m_\pi$, $\hat{O}$ begins at ${\cal O}(e^2)$ with the proton
Thomson amplitude $\sim \frac{-e^2}{M}$.
%$\frac{-e^2}{M} \varepsilon' \cdot \varepsilon$
%($\varepsilon$ and $\varepsilon'$ are photon polarization vectors).
At ${\cal O}(e^2 P)$ $\hat{O}$ includes ``pion-cloud'' mechanisms that
generate the leading contribution to the nucleon's electric and
magnetic polarizabilities, $\alpha$ and $\beta$. It also includes
analogous two-nucleon mechanisms where the incoming and outgoing
photons couple to what can be thought of as the deuteron's pion cloud,
i.e. the exchanged pions that generate the LO $\chi$PT
potential. These exchange currents are large, providing about 50\% of
the $\gamma$d dcs at $\omega=65$ MeV. At these energies the $\chi$ET
prediction is in good agreement with data (see left panel of
Fig.~\ref{fig:gammad}). However, the agreement with data at
95 MeV is not good (see right panel of Fig,~\ref{fig:gammad}).  Later
$\chi$ET calculations extended the calculation of $\hat{O}$ to ${\cal
  O}(e^2 P^2)$ and used $\chi$ET wave
functions~\cite{Beane05}. However, this leads to very little
improvement in the description of the 95 MeV data.

The rapid rise in the $\gamma$d dcs at backward angles is now
understood to be due to M1 excitation of the
Delta(1232) resonance~\cite{Hildebrandt2005}. Chiral EFTs
that include this resonance as an explicit degree of freedom
describe the backward-angle 95 MeV data well. 
%, and can also
%describe $\gamma$p data to higher energies than is possible with
%Delta-less $\chi$PT~\cite{ComptonDelta}. 
At the same time the
power counting in $\chi$ET for $\omega \sim \frac{m_\pi^2}{M}$
has been worked out. In this domain additional diagrams that 
ensure that the low-energy theorem for $\gamma$d
at $\omega=0$ is obeyed must be included, and these have now been
computed~\cite{Hildebrandt2006}. But these diagrams are formally 
and numerically sub-leading
for $\omega \sim m_\pi$. If they are included in the
computation of $\hat{O}$ for $\omega \approx 90$ MeV the variation in
the cross section due to short-distance physics in the $NN$ system is
reduced to 1--2\%~\cite{Hildebrandt2006}.

Therefore the elements needed for a $\chi$ET calculation of the
$\gamma$d dcs in the range $\omega=50$--100 MeV with an accuracy $\sim
3$\% are now understood. This ability of $\chi$ET to calculate the
$NN$ dynamics in such a controlled way motivates experimental efforts
that aim to use new $\gamma$d data to extract the isoscalar
combinations of nucleon electric and magnetic polarizabilities
$\alpha_N \equiv (\alpha_p + \alpha_n)/2$ and $\beta_N \equiv (\beta_p
+ \beta_n)/2$. One such experiment is underway at the MAX-Lab facility
at Lund, and will significantly increase the world data-base on the
$\gamma$d reaction~\cite{Maxlab}.  When this new data is used in
concert with a new, precision $\chi$ET calculation of $\gamma$d it
should yield an extraction of $\alpha_N - \beta_N$ with an accuracy
comparable to that with which $\alpha_p - \beta_p$ is presently known.
This will provide important constraints on the interplay between the
pion-cloud mechanisms that generate the dominant piece of the
nucleon's electric polarizability and other mechanisms that contribute
to $\alpha_N$ and $\beta_N$.

Recently it has been pointed out that elastic Compton scattering from
the Helium-3 nucleus also provides access to information on neutron
polarizabilities~\cite{Choudhury2007}. In this case the presence of
two protons in the nucleus significantly enhances the Compton cross
section. It also enhances (in absolute terms) the impact of $\alpha_n$
and $\beta_n$ on observables, because in coherent $\gamma {}^3$He
scattering the polarizability effects in the single-nucleon Compton
amplitude interfere with {\it two} proton Thomson terms.

\begin{figure}[ht]
\begin{center}
\includegraphics[width=4in]{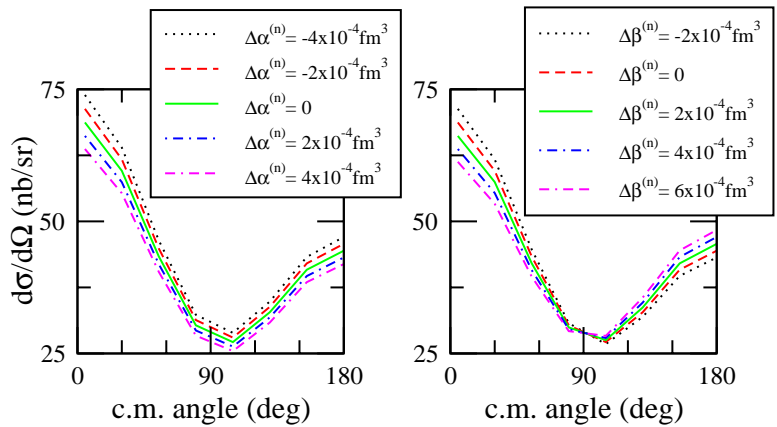}
\caption{Centre-of-mass frame differential cross section for $\gamma
  {}^3$He scattering at 80 MeV as predicted at $O(e^2 P)$, with the
  addition of shifts in $\alpha_n$ (left panel) and $\beta_n$ (right
  panel). Taken from Ref.~\cite{Choudhury2007}.}
\label{fig:gammaHe3}
\vspace*{-0.7cm}
\end{center}
\end{figure}

We have performed $\chi$ET calculations of $\gamma {}^3$He scattering
at ${\cal O}(e^2 P)$ [NLO] \cite{Choudhury2007}. These calculations
employ the same operator $\hat{O}$ as was used for $\gamma$d
scattering in Ref.~\cite{Beane99}, as well as a variety of $\chi$ET
three-nucleon wave functions that are consistent with $\hat{O}$ at
this order in $\chi$ET. This is the first consistent $\chi$ET
calculation of an electromagnetic reaction on the three-body system
(but see also Ref.~\cite{Song2007} for static electromagnetic
properties of the trinucleons) and---so far as I am aware---the first
calculation of $\gamma {}^3$He scattering.  Once
again, $\chi$ET makes a prediction for Compton observables at this
order, with the impact of $\alpha_n$ and $\beta_n$ on the dcs shown in
Fig.~\ref{fig:gammaHe3}.

We can also predict the asymmetries that would be obtained were
circularly polarized photons to scatter from Helium-3 nuclei polarized
parallel ($\Sigma_z$) or perpendicular ($\Sigma_x$) to the incoming
beam.  In the Helium-3 nucleus, the two protons are predominantly in a
${}^1$S$_0$ state, and so the double-polarization observables are
dominated by the contribution from the neutron that is (mostly)
carrying the spin of the polarized Helium-3.  For photon energies
above 100 MeV we find significant
sensitivity in $\Sigma_z$ and $\Sigma_x$ to neutron spin
polarizabilities~\cite{Ragusa}.
%Since there are four of these quantities and two
%are at present completely unconstrained experiments to measure these
Experiments that will measure these asymmetries are planned for the
HI$\gamma$S facility at the Triangle Universities Nuclear
Laboratory~\cite{HIGS}, and will provide important new constraints on
low-energy neutron spin structure.

\section{Other reactions on light nuclei in $\chi$ET: briefly}

\label{sec-other}

The dynamics of mesons and nucleons that is the focus of these
meetings has significant consequences for nuclear physics. In
particular, the $\pi$N interactions that are encoded in $\chi$PT are
now being used as the basis for a quantitative understanding of
few-nucleon systems. In this talk I have discussed only the portion of
this understanding that pertains to electromagnetic reactions. But the
$\chi$ET approach to nuclear dynamics has also had significant success
in describing both elastic scattering and breakup reactions in
neutron-deuteron and proton-deuteron experiments~\cite{ThreeN}. 
For comprehensive reviews of the application of $\chi$ET to few-nucleon
systems see Refs.~\cite{reviews}.

Meanwhile, $\chi$ET has also been used with significant success in
understanding low-energy weak processes~\cite{Park2002}. The fact that
the chiral Lagrangian provides a connection between these reactions
and pionic processes such as $\pi^- d \rightarrow nn
\gamma$~\cite{GardestigPhillips2006} is now being exploited to yield
new, more precise calculations of the latter
process~\cite{GardestigPhillips2007}. 
%Indeed, pionic interactions
%with the two-nucleon system have been a rich field for investigation
%using $\chi$ET methods for some time (see,
%e.g. Ref.~\cite{Beane97,Beane04,Krebs04}). 
Several contributions to this conference described progress in pionic
reactions on deuterium~\cite{Pidpiprod}. The resulting calculations
are evidence of the power of a description which allows us to trace
the consequences of QCD's chiral symmetry and the pattern of its
breaking through into predictions for experiments involving light
nuclei.

\section*{Acknowledgments}
I thank the organizers of MENU2007 for a very stimulating and
efficiently run meeting. I am also grateful for enjoyable and
educative collaborations with those who worked with me on the
research described here.  Responsibility for any aberrant views and/or
omissions in this paper is, however, entirely mine. This work was
supported by DOE grant DE-FG02-93ER40756.

\end{document}